\documentclass[apj]{emulateapj}
\pdfoutput=1
\usepackage{hyperref}
\usepackage{amsmath,amstext}
\usepackage[T1]{fontenc}
\usepackage[figure,figure*]{hypcap}

\shorttitle{The Calcium I Resonance Line in Cool DZs}
\shortauthors{Blouin et al.}

\begin{document}

\submitted{Accepted for publication in The Astrophysical Journal}

\title{Line Profiles of the Calcium I Resonance Line in Cool Metal-polluted White Dwarfs}

\author{S. Blouin\altaffilmark{1}}
\author{N.~F. Allard\altaffilmark{2,3}}
\author{T. Leininger\altaffilmark{4}}
\author{F.~X. Gad\'ea\altaffilmark{4}}
\author{P. Dufour\altaffilmark{1}}

\altaffiltext{1}{D\'epartement de Physique, Universit\'e de Montr\'eal, Montr\'eal, 
QC H3C 3J7, Canada; sblouin@astro.umontreal.ca, dufourpa@astro.umontreal.ca}
\altaffiltext{2}{GEPI, Observatoire de Paris, Universit\'e PSL, CNRS, UMR 8111,
  61 avenue de l'Observatoire, F-75014 Paris, France; nicole.allard@obspm.fr}
\altaffiltext{3}{Sorbonne Universit\'e, CNRS, UMR 7095,
  Institut d'Astrophysique de Paris, 98bis boulevard Arago, F-75014 Paris, France}
\altaffiltext{4}{Laboratoire de Physique et Chimie Quantique, UMR 5626, 
Universit\'e de Toulouse (UPS) and CNRS, 118 route de Narbonne,
F-31400 Toulouse, France}

\begin{abstract}
  Metal-polluted white dwarfs (DZ stars) are characterized by a helium-rich
  atmosphere contaminated by heavy elements traces originating from accreted
  rocky planetesimals. As a detailed spectroscopic analysis of those
  objects can reveal the composition of the accreted debris, there is a
  great interest in developing accurate DZ atmosphere models.
  However, the coolest DZ white dwarfs are challenging
  to model due to the fluidlike density of their atmospheres. Under
  such extreme conditions, spectral absorption lines are heavily
  broadened by interactions with neutral helium and it is no longer
  justified to use the conventional Lorentzian profiles.
  In this work, we determine the theoretical profiles of the \ion{Ca}{1}
  resonance line (the most prominent spectral line for the coolest
  DZ white dwarfs) in the dense atmospheres of cool DZ white dwarfs.
  To do so, we use a unified theory of collisional line profiles and
  accurate ab initio potential energies and transition dipole moments
  for the CaHe molecule.
  We present the resulting profiles for the full range of temperatures
  and helium densities relevant for the modeling of cool, metal-polluted white
  dwarfs (from 3000 to 6000\,K and from $10^{21}$ to $10^{23}$\,cm$^{-3}$).
  We also implement these new profiles in our atmosphere models and show
  that they lead to improved fits to the \ion{Ca}{1} resonance line of the coolest
  DZ white dwarfs.
\end{abstract}

\section{Introduction}
After exhausting their supplies of nuclear fuel, the vast majority of 
main-sequence stars will end their lives as white dwarfs. These stellar remnants 
are then condemned to a slow cooling that will extend over billions of years.
White dwarfs are compact objects ($\approx 0.6\,M_{\odot}$ compressed
into a volume similar to that of the Earth) and thus, they are characterized
by a very intense surface gravity. Because of this strong gravitational field
(typically, $\log g=8$), atomic species stratify according to their weights.
Hence, only the lightest elements---hydrogen and helium---are found in the
atmospheres of most white dwarfs.

A particularly interesting exception to this rule is the existence of DZ stars,
white dwarfs whose helium-rich atmospheres are contaminated by traces of heavy elements 
(e.g., Mg, Ca, Na, Fe, Si) that are detected thanks to their absorption lines.
The existence of those objects is a priori incompatible with
the very efficient gravitational settling at work in white dwarfs. In particular,
these heavy elements are expected to sink below the photosphere within timescales
that are much smaller than the age of the white dwarf 
\citep{paquette1986diffusion,koester2009accretion}.

The accepted scenario for the existence of DZ white dwarfs is that heavy
elements were recently (or are being) added to the atmospheres of those objects
through the accretion of rocky planetesimals 
\citep[e.g., see the reviews of][]{jura2014extrasolar,farihi2016circumstellar}. This scenario is 
now supported by the detection of infrared excesses that signal the presence of 
circumstellar debris disks around dozens of white dwarfs 
\citep{melis2010echoes,brinkworth2012spitzer,rocchetto2015frequency}
and the discovery of planetary transits in the light curve of WD~1145+017
\citep{vanderburg2015disintegrating}.
Metal-polluted white dwarfs constitute a unique observational window into the
bulk composition of planetesimals (comets, asteroids, dwarf planets) outside
our Solar System. Using atmosphere models to fit the observed spectra of DZ
white dwarfs, we can determine the chemical composition of their atmospheres and
trace back the composition of the accreted planetesimals
\citep{zuckerman2007chemical,farihi2013evidence,hollands2018cool}.

For the coolest ($T_{\rm eff} \approx 4000\,{\rm K}$)---and thus, oldest---white dwarfs, very few metal spectral lines remain visible. In this regime,
one of the most prominent lines is the \ion{Ca}{1} resonance line at 4226\,{\AA}
\citep[see for instance WD~2251$-$070 in][]{dufour2007spectral}.
Properly modeling this absorption feature is of utmost importance if we want
to precisely determine the composition of these old objects.
For instance, one of the oldest DZ white dwarf known to date (WD~J2356$-$209, with a
cooling age of $\approx 8\,$Gyr) has
been shown to have an abnormally high Na/Ca abundance ratio, which raises a number
of questions about the origin of the planetesimal that it accreted \citep{blouin2019WD2356}.
The Ca abundance in the atmosphere of this object was inferred from a fit to its
\ion{Ca}{1} resonance line. Therefore, the determination of the atmospheric composition of
this white dwarf and any conclusions about the nature of the accreted planetesimal
are strongly influenced by the quality of our \ion{Ca}{1} 4226\,{\AA} line profiles.

However, obtaining accurate absorption line profiles under the physical conditions found at
the photosphere of those cool, helium-rich white dwarfs is challenging. Because of their very
transparent atmospheres, cool DZ white dwarfs have photospheres where 
the density can reach $n_{\rm He} = 10^{23}\,{\rm cm}^{-3}$ \citep{blouin2018dztheory}. 
Under such fluid-like conditions, the wings of heavy-element spectral lines are
strongly broadened by interactions with neutral helium and Lorentzian profiles do not
allow a satisfactory fit to the observed spectra \citep[e.g.,][]{allard2016asymmetry}. 
It is then necessary to implement
a unified line shape theory \citep{allard1999effect} to obtain the right spectral line profiles.
This approach requires prior knowledge, for each molecular state involved in the transition, 
of the potential energy and the variation of the dipole moment with respect to the 
atom-atom separation.

The accuracy of the line profiles is strongly affected by the quality of the atomic
data from which they are computed. The study of SDSS J080440.63+223948.6 (SDSS~J0804+2239) by
\cite{blouin2018dzcia} offers a good example of the importance of using high-quality
ab initio data for these calculations. As no accurate potential energies for the
CaHe molecule were available, \cite{blouin2018dzcia} tried to model the \ion{Ca}{1} resonance line of 
SDSS~J0804+2239 using approximate potential energy curves (PECs). The limitations
of their approximate potentials were quite obvious, as they were unable to obtain
a completely satisfactory fit to this absorption line.

For their analysis of WD~J2356$-$209, \cite{blouin2019WD2356} relied on more accurate
\ion{Ca}{1} 4226\,{\AA} line profiles---thanks to the improved PECs of
\citealt{hernando2008absorption} (M.~Barranco 2018, private communication)---and achieved an excellent fit to the \ion{Ca}{1} resonance line of this object.
Still, \cite{blouin2019WD2356} did not have access to the transition dipole moments and 
to the long-range portion of the PECs, which are both needed to obtain the most
accurate line profiles.
In this paper, we use state-of-the-art ab initio data to fill
these gaps and obtain more accurate \ion{Ca}{1} resonance line profiles. 
The new ab initio data on which this work relies are described in Section \ref{sec:abinitio}.
In Section \ref{sec:results}, we present our new \ion{Ca}{1} resonance line profiles.
We apply these new profiles to three cool DZ white dwarfs
in Section \ref{sec:applications} and our conclusions are given in Section
\ref{sec:conclusion}.

\section{Ab initio data}
\label{sec:abinitio}
Line profile intensities are functions of both excited and ground-state interactions.
Therefore, a precise determination of the electronic energies and optical
transition dipole moments is crucial to accurately compute the \ion{Ca}{1} resonance line
profiles for the whole wavelength range.
To obtain the necessary atomic data, we performed ab initio calculations with
the MOLPRO package \citep{molpro}.
All singlet states up to the Ca $^{1}$S (4s5s) + He (1s$^2$)
asymptote have been involved (i.e., four $\Sigma$ states, two $\Pi$ states and one
$\Delta$ state). A large core pseudopotential was used
for Ca \citep{czuchaj1991pseudopotential}
together with the usual Core Polarisation Potential
approach \citep{muller1984treatment}.
For Ca, the associated 8s8p7d basis set of \citeauthor{czuchaj1991pseudopotential} has been complemented by 7f and 3g gaussian functions, while for He, the spdfg
aug-v5z-cc basis set has been used \citep{woon1984gaussian}. State-averaged 
CASSCF (complete active space self-consistent field) calculations
with four active electrons in eleven orbitals (corresponding
to the 4s, 4p, 3d, 5s orbitals for Ca and the 1s orbital for He)
were performed, followed by MRCI (multireference configuration interaction) 
calculations for each symmetry. The
transition dipole moments were then computed using the averaged natural
orbitals from the bra wavefunction.

\begin{figure}
    \includegraphics[width=\columnwidth]{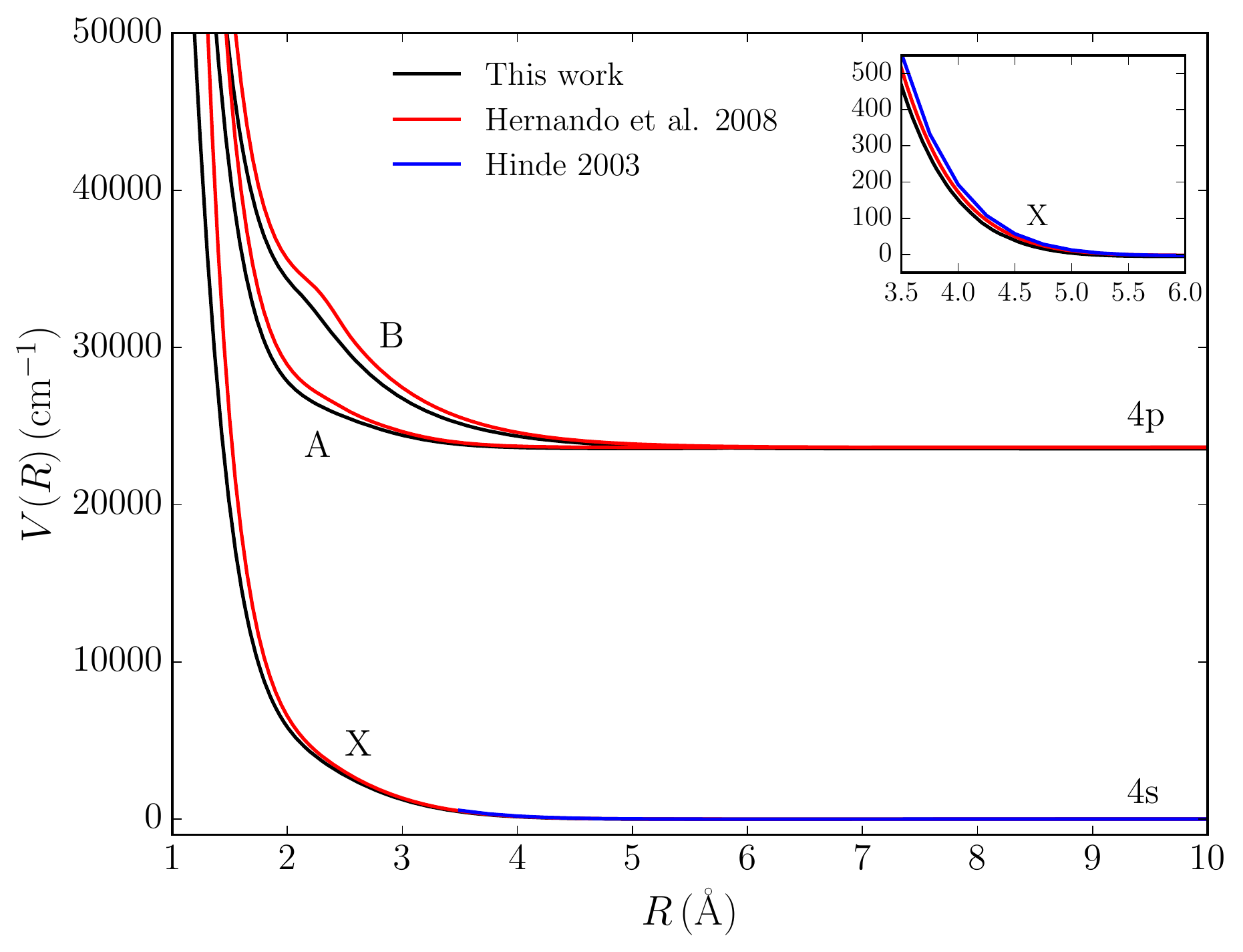}
    \caption{PECs for the 4s and 4p states of the CaHe
      molecule. For comparison, the potentials computed by
      \cite{hinde2003mg} and \citealt{hernando2008absorption}
      (which were used in \citealt{blouin2019WD2356})
      are also shown.}
  \label{fig:potCaHe}
\end{figure}

The resulting potentials are reported in Figure \ref{fig:potCaHe}. The ground state
(X 1\,$^{1}\Sigma$) is mainly repulsive with a shallow van der Waals
well, the 2\,$^{1}\Sigma$ and 2\,$^{1}\Pi$ (A) potentials are more
attractive and similar to the ground-state potential of the Ca$^+$He ion, and
the 3\,$^{1}\Sigma$ (B) potential is very repulsive. From the
generalized Fermi model \citep{dickinson2002undulations}, because the electron$-$He
scattering length is positive, repulsive effects are expected. Moreover,
since the $^{1}$D (4s3d) and $^{1}$P (4s4p) asymptotes are rather
close in energy, it is not surprising that the highest B state (3\,$^{1}\text{\ensuremath{\Sigma}}$)
cumulates the repulsive interactions with He, while the lower one
does not and looks like the ground-state potential of the Ca$^+$He ion. 

\section{Line profiles}
\label{sec:results}
To compute the pressure broadening of the \ion{Ca}{1} resonance line, we rely on the classical path
expression derived by \cite{allard1999effect}. In contrast with other more usual
approximations \citep{anderson1952method,baranger1958simplified,baranger1958problem},
the approach of \citeauthor{allard1999effect} takes into account the
dependence of the electric dipole moment on the position of perturbers, which can 
significantly affect the spectral line shape \citep{allard1998new,allard1998satellites}.

\subsection{Low Densities}
\begin{figure}
    \includegraphics[width=\columnwidth]{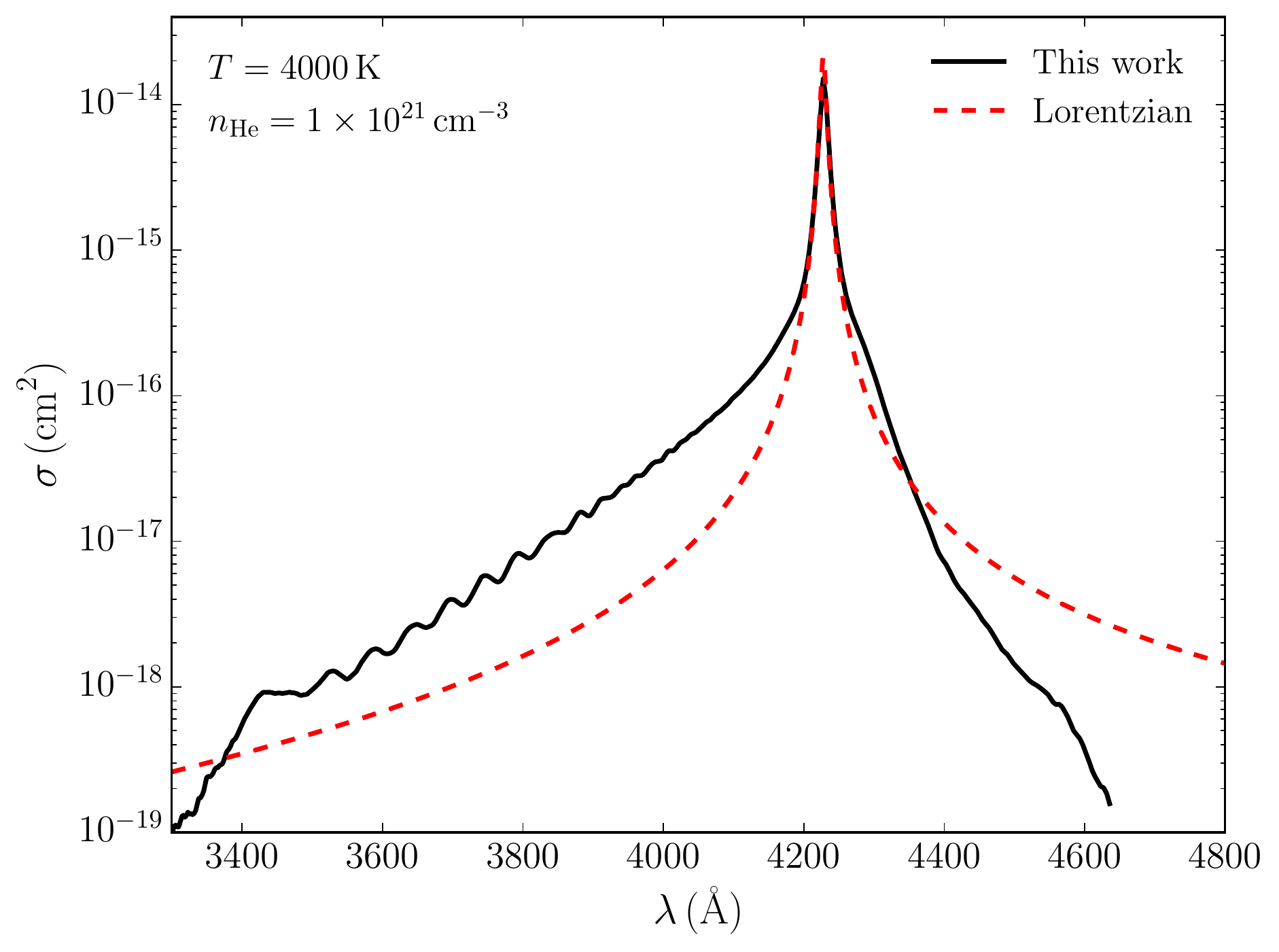}
    \caption{Comparison  of the  unified  profile (solid black  line)  with the corresponding Lorentzian profile (dashed red line).}
  \label{fig:comp_lorentz}
\end{figure}

\begin{figure}
    \includegraphics[width=\columnwidth]{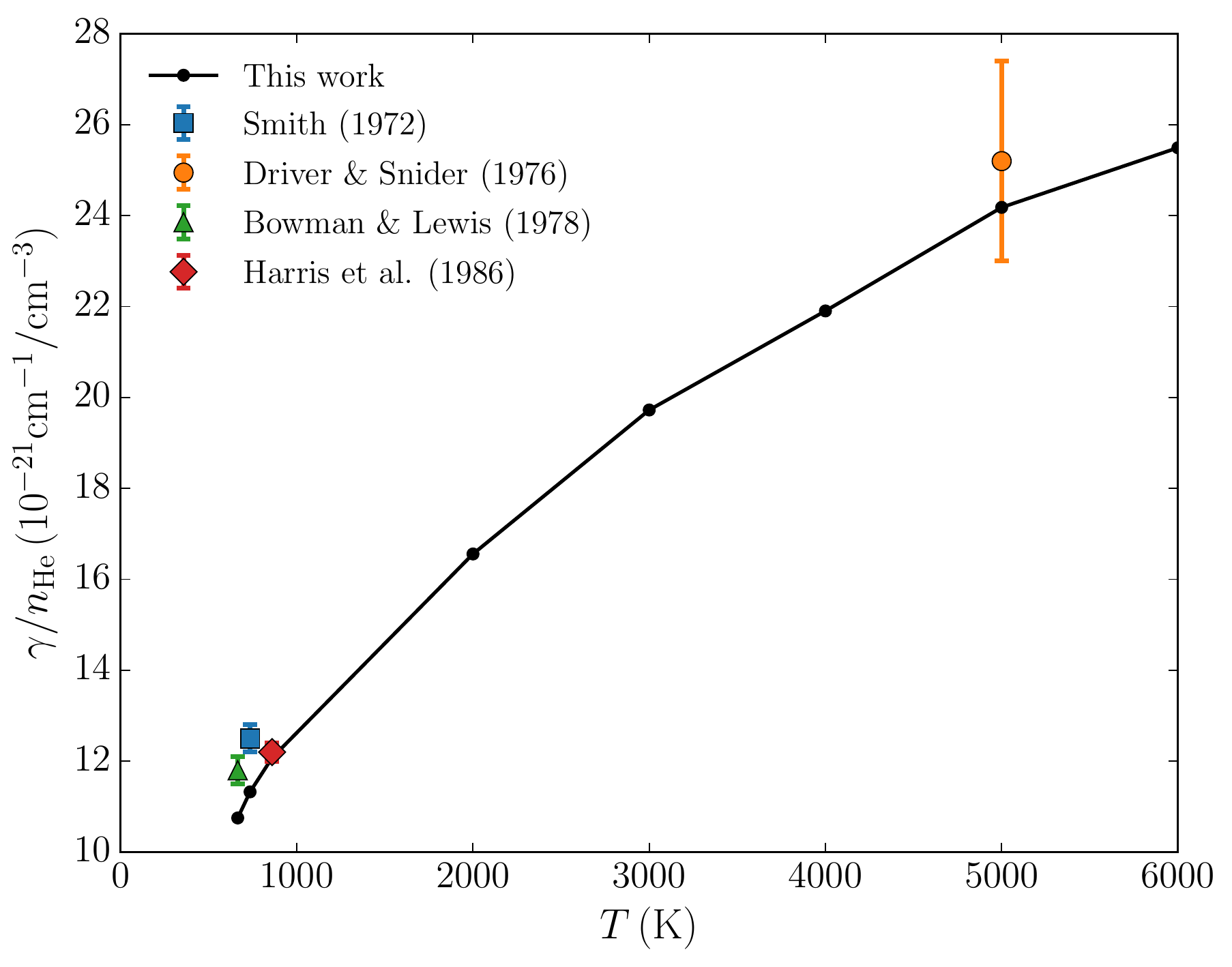}
    \caption{Variation with temperature of the half width at half maximum 
      of the \ion{Ca}{1} resonance line perturbed by He collisions. Our results
      are shown in black and the error bars represent experimental measurements.} 
    \label{fig:exp_comp}
\end{figure}

Under sufficiently low perturber densities, the center of the spectral line is expected to be symmetric
and described by a Lorentzian profile. This profile is defined by two parameters,
a width and a shift, which have a linear dependence on the density.
We find that the core of the line is adequately described by a Lorentzian
up to $n_{\rm He}=10^{21}\,{\rm cm}^{-3}$ (Figure \ref{fig:comp_lorentz}).
Using the impact limit of the general calculation of the autocorrelation function
\citep{allard1999effect}, we computed the line parameters that characterize this Lorentzian core.
As shown in Figure \ref{fig:exp_comp}, the resulting theoretical broadening parameters
are in reasonable agreement with laboratory measurements \citep{smith1972collision,driver1976broadening,bowman1978collision,harris1986measurement}. Note that the error bars on the experimental data are probably underestimated, since the three points at low temperature appear to imply an improbable temperature dependence.

While the core of the resonance line is adequately described by a Lorentzian profile
at $n_{\rm He}=10^{21}\,{\rm cm}^{-3}$, it is of course not the case for the wings
(Figure \ref{fig:comp_lorentz}).
The X$-$B transition is responsible for the shape of the blue wing and the X$-$A transition
affects the red side. The small maximum in the blue wing near 3400\,{\AA} is a direct
consequence of the shoulder of the B state PEC at $R=2.2\,$\AA, which
gives rise to a local maximum of $\Delta V (R)$ at $\Delta V=5800\,{\rm cm}^{-1}$.
As satellite lines occur at frequencies corresponding to extrema of $\Delta V$,
a satellite feature appears at $\Delta \omega = 5800\,{\rm cm}^{-1}$ (which
corresponds to $\lambda \simeq 3400\,$\AA).\newpage

\subsection{High Densities}
\begin{figure}
    \includegraphics[width=\columnwidth]{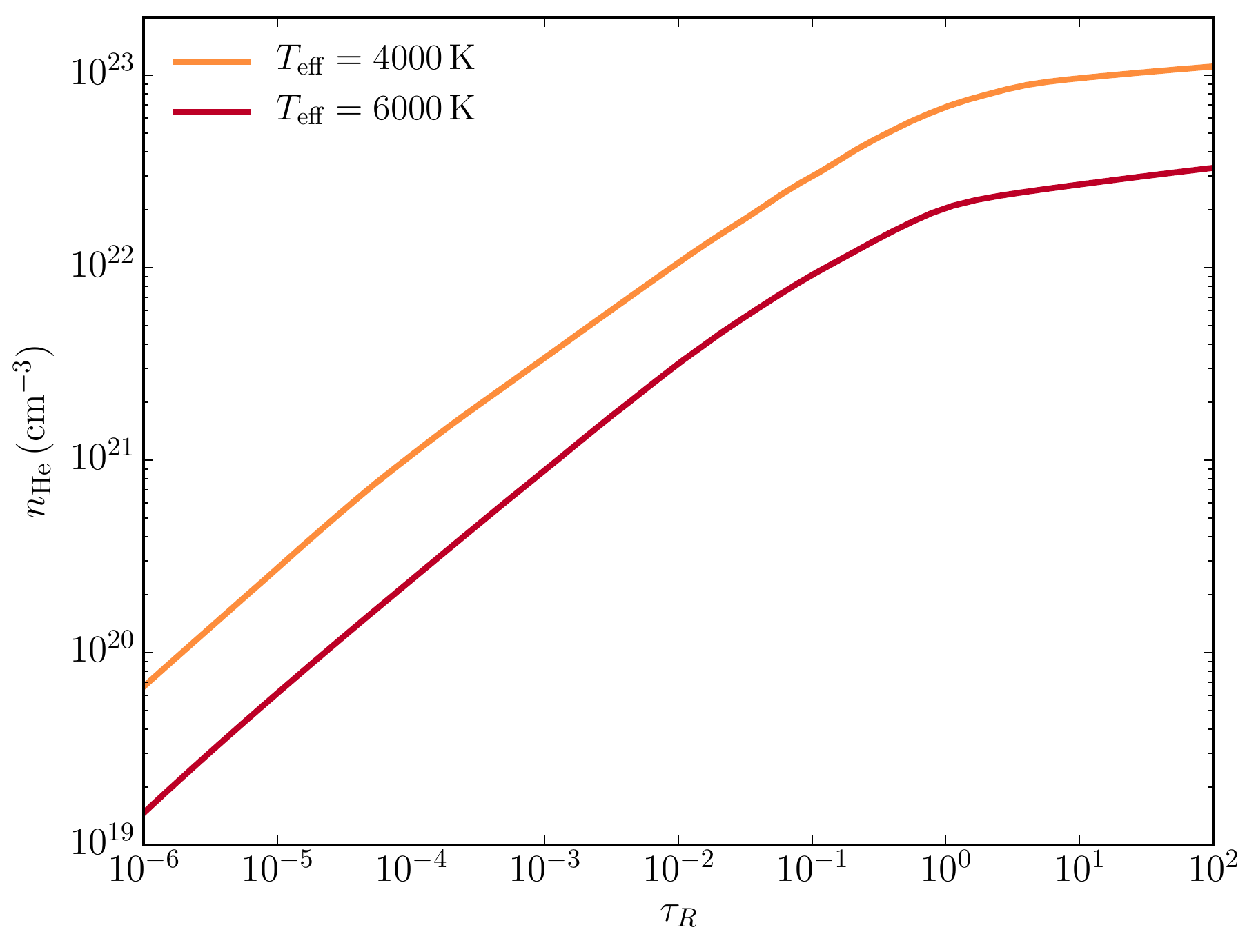}
    \caption{He density as a function of the Rosseland mean optical depth for
      two models of cool DZ atmospheres. Note that a hydrogen-free atmosphere
      with a calcium abundance $\log\,{\rm Ca/He}=-10$ and a surface gravity $\log g =8$
      was assumed for both models. The abundance ratios of all other heavy elements
      were scaled to the abundance of Ca to match the abundance ratios measured in
      CI chondrites \citep{lodders2003solar}.}
  \label{fig:structures}
\end{figure}

The coolest DZ white dwarfs are characterized by photospheric densities that can go up
to $n_{\rm He}=10^{23}\,{\rm cm}^{-3}$, with much of their line-forming regions located
at densities exceeding $10^{21}\,{\rm cm}^{-3}$ (Figure \ref{fig:structures}). 
As those cool objects are precisely the ones that show a prominent \ion{Ca}{1} resonance line, there is a strong
astrophysical interest in properly modeling its line shape under such high-density conditions.

Given the high perturber densities involved,
simultaneous collisions are very frequent. This implies that many-body collisions
must be included in our calculation of the \ion{Ca}{1} resonance line pressure broadening and this
is why we rely on the autocorrelation formalism of \cite{allard1999effect}.
This approach has already proved successful for the modeling of spectral lines
in cool DZ white dwarfs \citep[e.g.,][]{allard2016asymmetry,allard2018line}.
Another proof of the accuracy of the \cite{allard1999effect} formalism has
been provided by the study of He doped with alkali atoms, which has recently been
a subject of active study \citep[e.g.,][]{hernando2010absorption,mateo2011absorption}.
In such experiments, large He clusters produced in a supersonic jet are doped
with alkali atoms and characterized by laser-induced fluorescence.
In the case of Na atoms attached to He droplets \citep{stienkemeier1996spectroscopy},
\cite{allard2013absorption} have shown that calculations performed
within the autocorrelation framework can successfully explain the laboratory spectra.
Moreover, it was also shown that the absorption spectra obtained by this approach
are consistent with those obtained from path integral Monte Carlo simulations
and the Franck-Condon approximation \citep{nakayama2001path}.

\begin{figure}
    \includegraphics[width=\columnwidth]{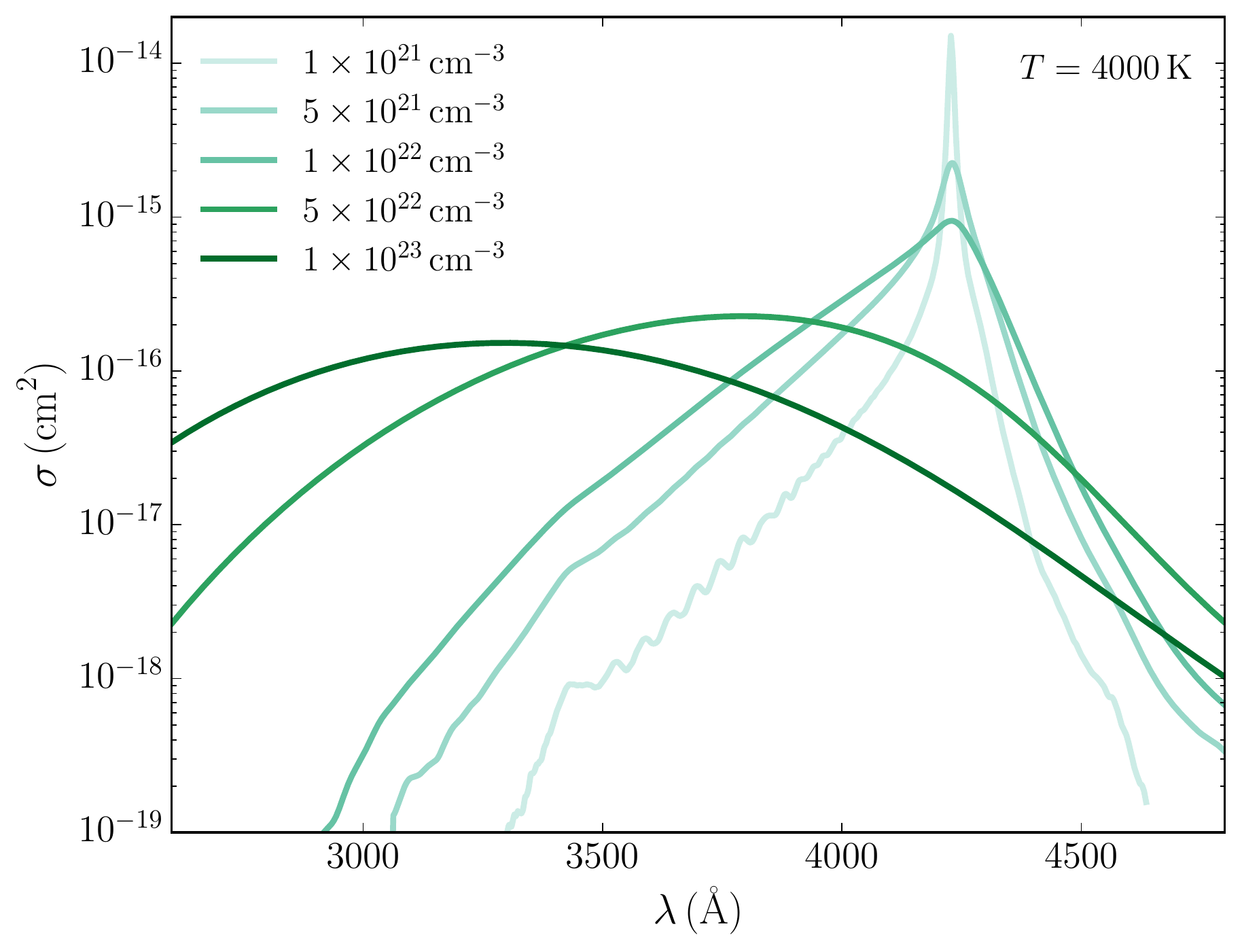}
    \caption{Density dependence of the \ion{Ca}{1} resonance line profiles. A constant
      temperature $T=4000\,{\rm K}$ is assumed.}
  \label{fig:dens_dep}
\end{figure}

\begin{figure}
    \includegraphics[width=\columnwidth]{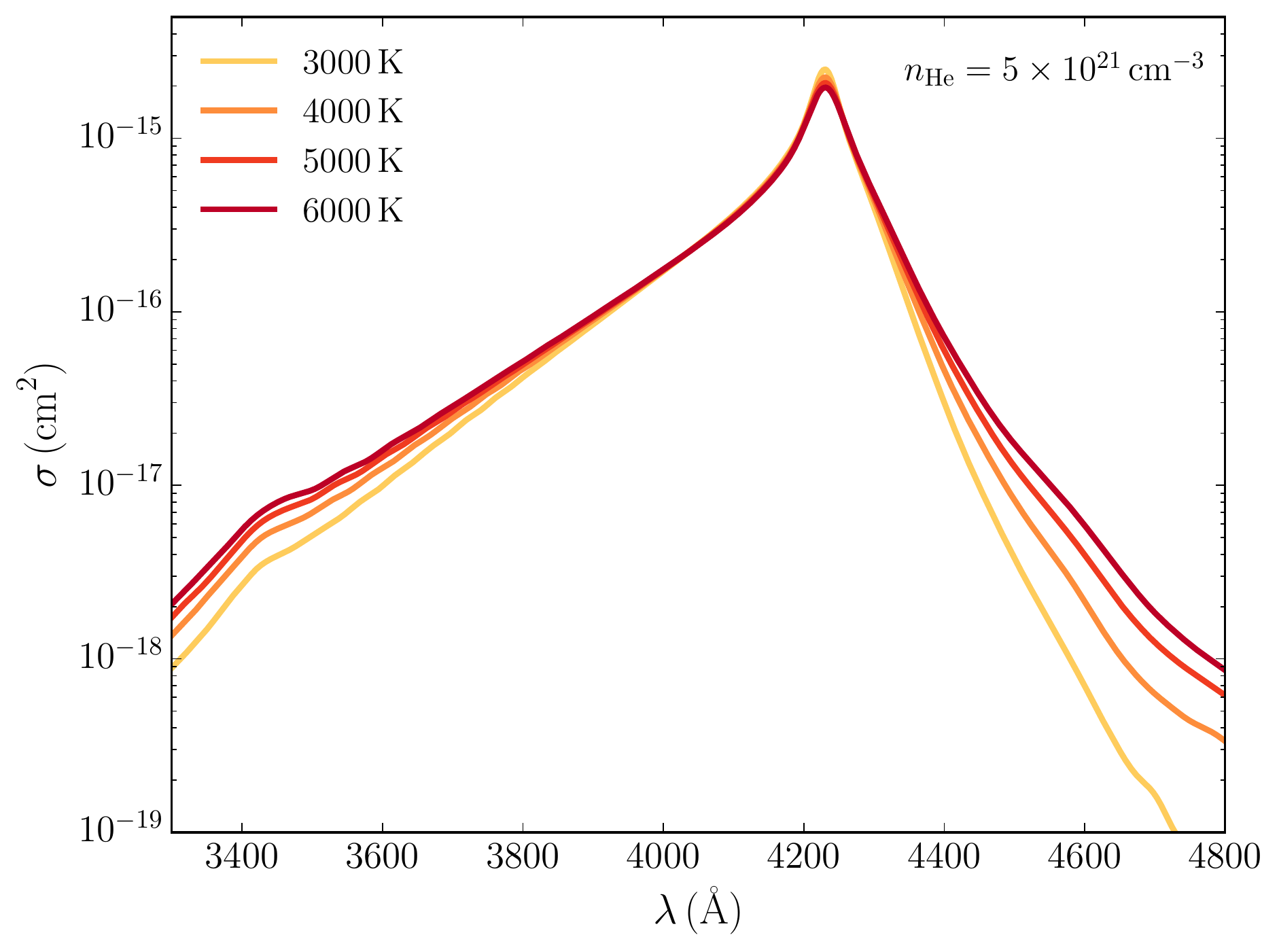}
    \caption{Temperature dependence of the \ion{Ca}{1} resonance line profiles. A constant
    density of $n_{\rm He}=5 \times 10^{21}\,{\rm cm}^{-3}$ is assumed.}
  \label{fig:temp_dep}
\end{figure}

Figure \ref{fig:dens_dep} shows the results of our calculations for
the evolution of the \ion{Ca}{1} resonance line profile with increasing He density. Clearly,
above $10^{21}\,{\rm cm}^{-3}$, the profiles become strongly asymmetric and are
shifted towards smaller wavelengths. Note that, as expected, the maximum of the
line shifts towards the satellite line at 3400\,{\AA}.
Apart from being strongly affected by the pertuber density, line profiles can also
be affected by the temperature of the medium \citep{allard2004temperature}.
However, for the temperature range relevant to cool DZ atmospheres, we find that
the profiles are not very sensitive to the temperature (Figure \ref{fig:temp_dep}).

\section{Astrophysical Applications}
\label{sec:applications}
\subsection{SDSS J0804+2239}
\begin{figure}
    \includegraphics[width=\columnwidth]{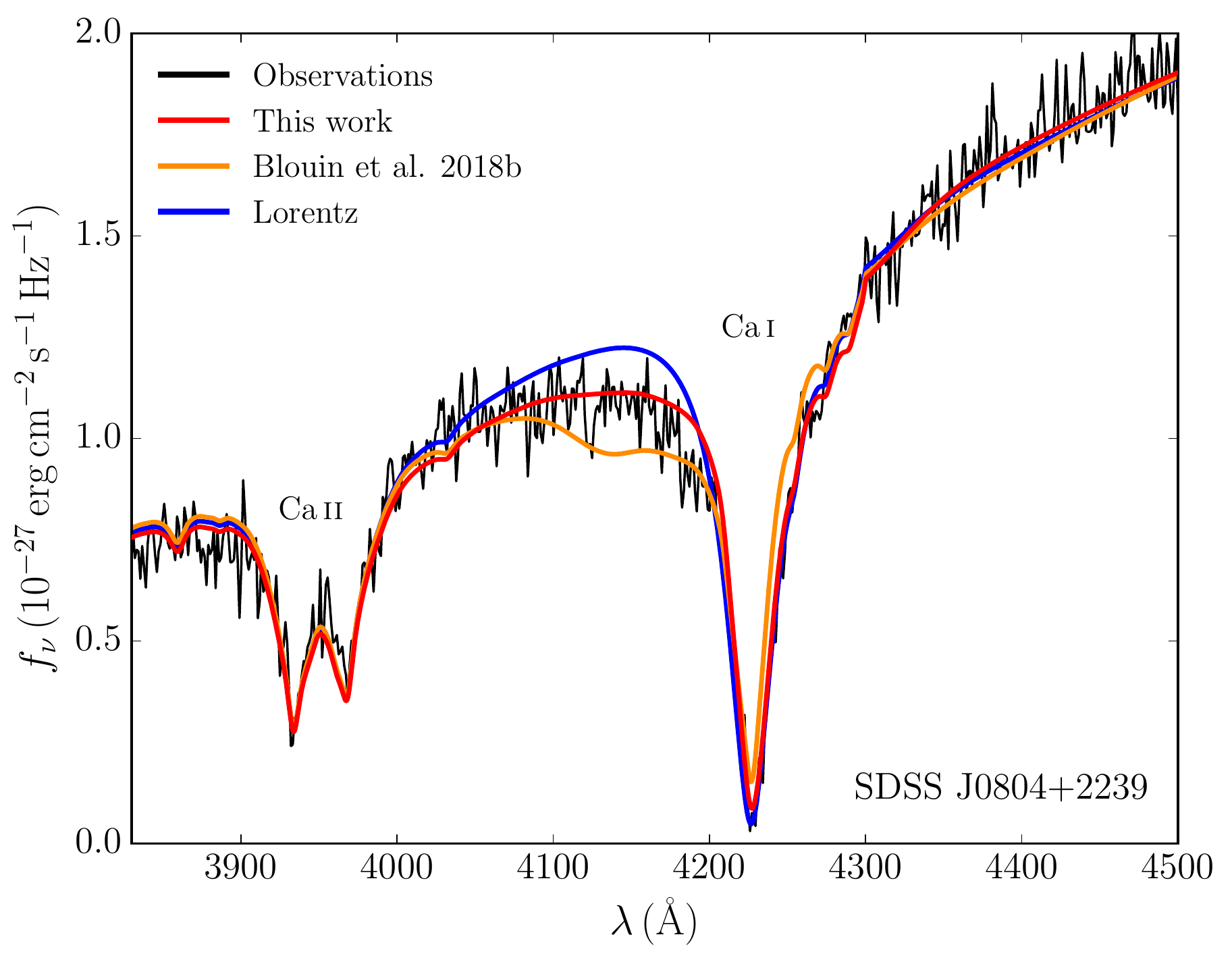}
    \caption{
      Comparison of synthetic spectra computed with different line profiles
      for the \ion{Ca}{1} 4226{\,\AA} transition. Observations of J0804+2239
      are in black, results found when assuming
      a Lorentzian profile are in blue and synthetic spectra obtained with the unified
      line shape theory of \cite{allard1999effect} are in orange for the
      case where approximate CaHe potentials were used \citep{blouin2018dzcia}
      and in red for the case where the high-quality CaHe potentials presented in this work
      were used. All synthetic spectra were
      computed assuming the atmospheric parameters of SDSS J0804+2239 found in
      \citet[i.e., $T_{\rm eff}=4970\,$K, $\log g=7.98$, $\log {\rm H/He}=-1.6$ and
        $\log {\rm Ca/He}=-10.0$]{blouin2018dzcia}.  Note that in all cases the
      \ion{Ca}{2} H \& K lines are computed with the profiles described in
      \cite{allard2014caii}.
    }
    \label{fig:J0804new}
\end{figure}

A comprehensive analysis of the DZ white dwarf SDSS~J0804+2239 was recently presented in 
\cite{blouin2018dzcia}. Thanks to the improved constitutive physics of their atmosphere
models, they were able to find a satisfactory fit to both its metal spectral lines and
its spectral energy distribution, which is affected by H$_2-$He collision-induced absorption.
They found an effective temperature of $T_{\rm eff}=4970\pm100\,{\rm K}$, a surface gravity of
$\log g = 7.98 \pm 0.05$ and number abundances of $\log\,{\rm H/He}=-1.6 \pm 0.2$ and
$\log\,{\rm Ca/He}=-10.0 \pm 0.1$. However, their spectroscopic fit of SDSS J0804+2239
was tarnished by their inability to adequately fit the blue wing of the \ion{Ca}{1} resonance line,
which they attributed to the poor quality of the CaHe PECs used to compute the profiles
of this spectral line.
Using the same atmospheric parameters
($T_{\rm eff}$, $\log g$ and individual abundances)
as those found by \cite{blouin2018dzcia}, we fitted
the Ca lines of SDSS~J0804+2239 with our improved \ion{Ca}{1} 4226\,{\AA} profiles. As shown in 
Figure~\ref{fig:J0804new}, these new profiles allow us to obtain an improved spectroscopic fit to 
the blue wing of the \ion{Ca}{1} resonance, solving the problem identified by \cite{blouin2018dzcia}.

\subsection{WD~J2356$-$209}
\begin{figure}
    \includegraphics[width=\columnwidth]{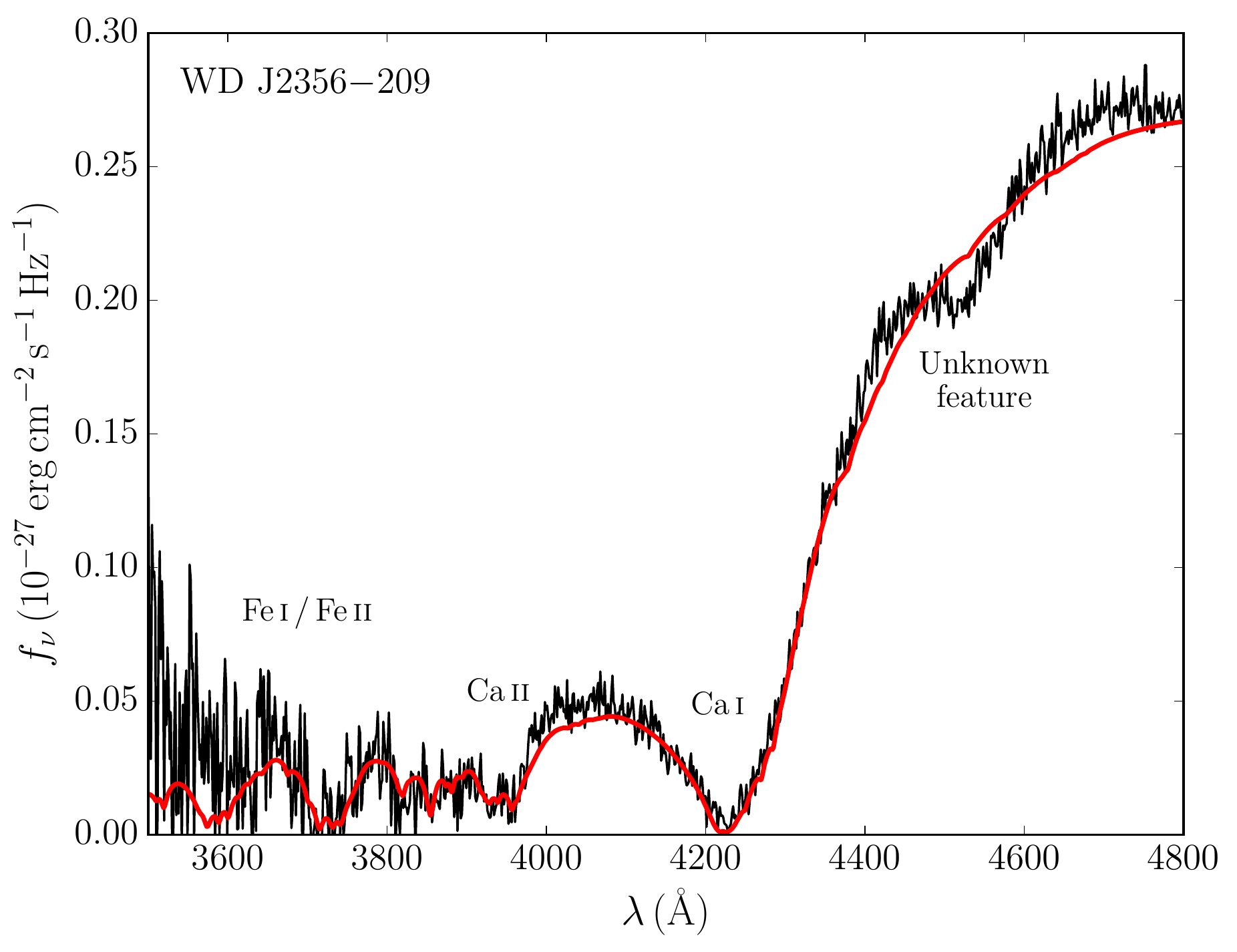}
    \caption{Spectroscopic fit of the \ion{Ca}{1} resonance line of WD~J2356$-$209. As in
    \cite{blouin2019WD2356}, $T_{\rm eff}=4040\,{\rm K}$, $\log g = 7.98$ and
    $\log\,{\rm Na/He}=-8.3$.
    However, the calcium abundance is now $\log\,{\rm Ca/He}=-9.4 \pm 0.1$,
    which is slightly lower than the value reported by \cite{blouin2019WD2356}.}
  \label{fig:2356fit}
\end{figure}

WD~J2356$-$209 is a very cool DZ white dwarf whose visible spectrum is dominated by
a very broad sodium feature \citep{salim2004cool}. 
\cite{blouin2019WD2356} have obtained a fit that is in
very good agreement with observations across all wavelengths and they have shown that
WD~J2356$-$209 has a record Na/Ca abundance ratio.
However, some uncertainty persisted as the \ion{Ca}{1} resonance profiles were computed assuming
constant transition dipole moments and as the long-range portion of the CaHe PECs was not
available. Thanks to our new ab initio data on the CaHe molecule,
the uncertainties that could potentially plague the Ca abundance determination are no longer present.
Using our \ion{Ca}{1} 4226\,{\AA} profiles, we performed a new spectroscopic fit of WD~J2356$-$209 
(Figure \ref{fig:2356fit}) and we found virtually the same abundances.
In particular, we found a calcium abundance of $\log\,{\rm Ca/He}=-9.4 \pm 0.1$, 
which is within the uncertainties of the $-9.3 \pm 0.1$ value of \cite{blouin2019WD2356}. Therefore, their 
conclusions are unchanged despite the addition of our new profiles: 
WD~J2356$-$209 does have a uniquely high Na/Ca abundance ratio. We note, however, that our new
profiles cannot account for the absorption feature near 4500\,{\AA}. The nature of this
absorption feature remains unknown
(see Section 4.3 of \citealt{blouin2019WD2356} for a comprehensive discussion on this issue).

\subsection{WD~2251$-$070}
After WD~J2356$-$209, WD~2251$-$070 is the second coolest known DZ white dwarf.
Its spectrum is quite different from that of WD~J2356$-$209, as it has a prominent
\ion{Ca}{1} resonance line but does not show a strong Na~D doublet \citep[][Figure 5]{blouin2019WD2356}.
While reasonable fits to its spectrum have been
obtained in previous studies \citep{kapranidis1986chemical,dufour2007spectral},
atmosphere model codes used to analyze this object relied on constitutive physics
that are not totally appropriate for the dense atmosphere of this cool white dwarf.

In particular, the analysis of \cite{kapranidis1986chemical} relied on models
that assumed a Thomas-Fermi equation of state \citep{kapranidis1983model}. Because of
this assumption, they found that electron thermal conduction was an important
energy transport mechanism, a conclusion that does not hold if we use more realistic
equations of state \citep{bergeron1995new,kowalski2007equation}. Also,
\cite{kapranidis1986chemical} computed their synthetic spectra from pure helium
atmosphere structures, thus ignoring the impact of heavy elements on the pressure
and temperature stratification. Moreover, a simple Lorentzian profile was assumed
for the \ion{Ca}{1} resonance line. Despite all those simplifying assumptions, they
managed to obtain an excellent fit to the resonance line
(with $T_{\rm eff} = 4500\,{\rm K}$ and $\log\,{\rm Ca/He}=-6.3$)
and were even able
to find a good agreement to the unknown feature at 4500\,{\AA}. Unfortunately,
no detail regarding this puzzling structure was given so we cannot know how
they were able to explain it.

\cite{dufour2007spectral} revisited WD~2251$-$070 using more realistic atmosphere
models, where heavy elements are included in the calculation of the atmosphere
structures. That being said, a number of unjustified approximations remained,
such as the ideal gas law, the ideal Saha ionization equilibrium and Lorentzian
profiles. Their best solution ($T_{\rm eff} = 4000\,{\rm K}$ and
$\log\,{\rm Ca/He}=-10.5$) was in good agreement with the photometry and with the
\ion{Ca}{1} resonance line (but they were unable to explain the absorption feature near 4500\,{\AA}).

\begin{figure}
    \includegraphics[width=\columnwidth]{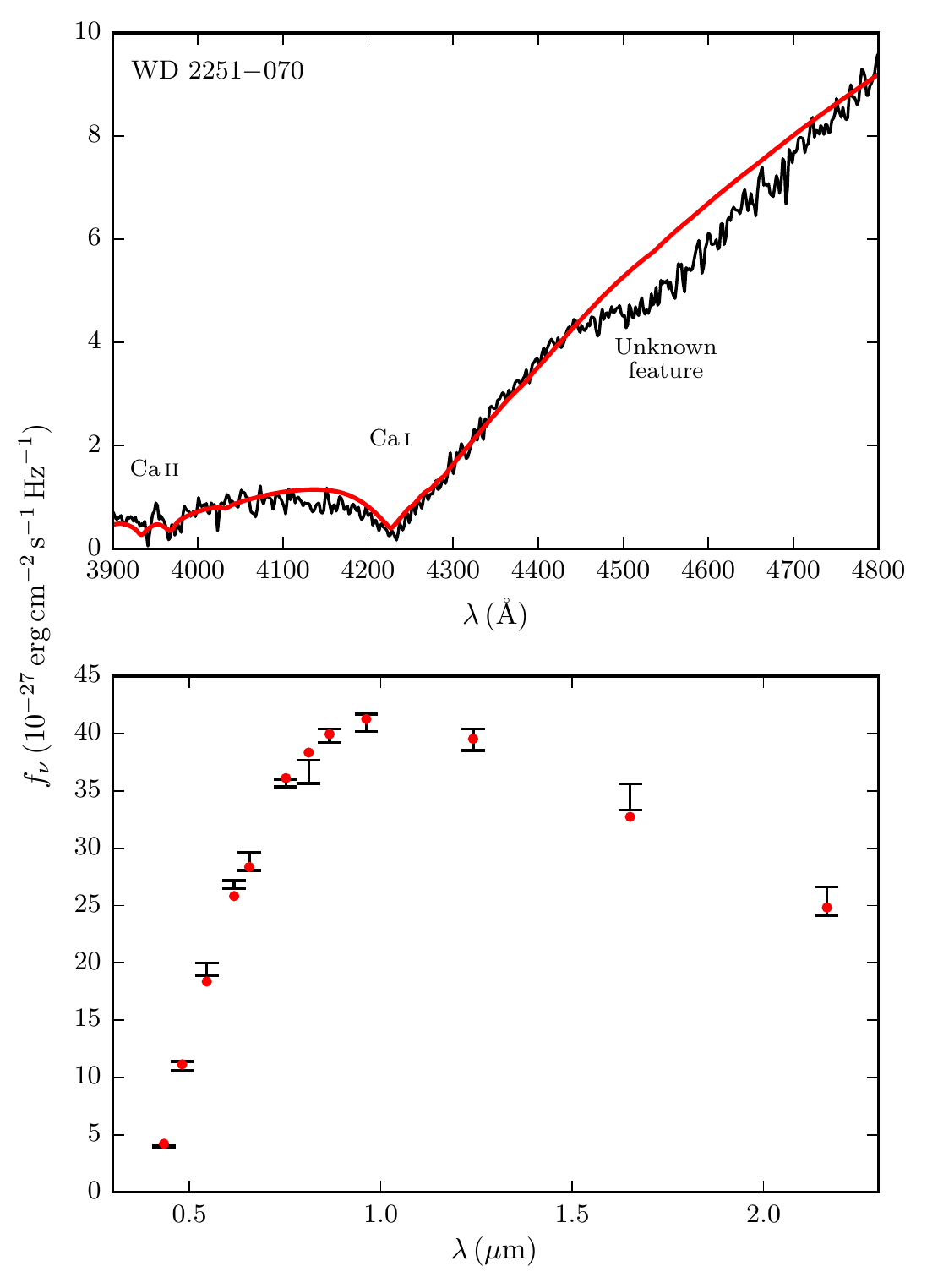}
    \caption{Our best solution for WD~2251$-$070. The top panel shows our
      spectroscopic fit
      and the bottom panel shows our fit to the $BVRI$, $grizy$ and
      $JHK$ photometry.
    }
  \label{fig:2251fit}
\end{figure}

Here, we revisit this star using our improved models \citep{blouin2018dztheory}
and our new \ion{Ca}{1} resonance line profiles. To do so, we use the spectroscopic observations of
\cite{dufour2007spectral}, the \textit{Gaia} data release 2 (DR2)
parallax measurement
\citep[$\pi = 117.15 \pm 0.05\,{\rm mas}$,][]{prusti2016gaia,brown2018gaia},
and photometry from \cite{bergeron1997chemical},
the Panoramic Survey Telescope And Rapid Response System
\citep[Pan-STARRS,][]{chambers2016panstarrs},
and the Two Micron All-Sky Survey (2MASS).
Our best fit to the spectroscopic and photometric observations is shown
in Figure \ref{fig:2251fit} and the corresponding atmospheric parameters
are given in Table \ref{tab:2251}.
 
\begin{deluxetable}{cc}
  \tablecaption{WD~2251$-$070 atmospheric parameters \label{tab:2251}}
  \tablehead{\colhead{\hspace{0.7cm}Parameter}\hspace{0.7cm} &
    \colhead{\hspace{0.7cm}Value}\hspace{0.7cm}}
  \startdata
  $T_{\rm eff}$       & $4170 \pm 90\,{\rm K}$\\
  $\log g$            & \hspace{0.08em}$8.06 \pm 0.08$\\
  $\log\,{\rm H/He}$  & $<-4.5$\\
  $\log\,{\rm Ca/He}$ & $-9.8 \pm 0.2$\hspace{0.75em}
  \enddata
\end{deluxetable}

To obtain this solution, we used the photometric technique
\citep{bergeron1997chemical} to determine $T_{\rm eff}$ and $\log g$.
More precisely, the solid angle $\pi (R/D)^2$ and the effective
temperature were found by adjusting the model fluxes to the photometric
observations. From the solid angle, we computed the white dwarf
radius $R$ ($D$ is given by the \textit{Gaia} parallax) and, from there,
we found the mass and the surface gravity using the evolutionary
models of \cite{fontaine2001potential}.
Using the parameters given by the photometric fit, we then adjusted
the Ca abundance to the spectroscopy. As the abundance obtained from
this spectroscopic fit was different from that originally assumed, we
repeated the whole fitting procedure (the photometric and the
spectroscopic fits) until we reached a consistent solution. Additionally,
as no collision-induced absorption is visible in the infrared photometry,
we were able to constrain the hydrogen abundance to $\log\,{\rm H/He}<-4.5$.

As shown in Figure \ref{fig:2251fit}, we are able to find a good fit
to most of the spectrum of WD~2251$-$070
(with the notable exception of the unknown absorption feature near 4500\,{\AA}).
In particular, our best solution is in good agreement with the observations
in the region located between the
\ion{Ca}{2} H \& K doublet and the \ion{Ca}{1} resonance line, whereas the best
solution of \cite{dufour2007spectral} was significantly above the observed
spectrum in this region. Moreover, unlike \cite{kapranidis1986chemical}, we
find a reasonable match to the weak \ion{Ca}{2} H \& K doublet.

\section{Conclusion}
\label{sec:conclusion}
\ion{Ca}{1} resonance line profiles suitable for the extreme densities
of the atmospheres of cool, metal-polluted white dwarfs were presented.
These new profiles were computed using state-of-the-art ab initio data and
a unified theory of collisional line profiles. At low densities, they
are compatible with laboratory measurements and, at high densities, they
are heavily broadened and shifted towards lower wavelengths.
We implemented these new profiles in our model atmosphere code and showed
how they lead to good spectroscopic fits for three cool DZ white dwarfs
(SDSS~J0804+2239, WD~J2356$-$209 and WD~2251$-$070).
However, in the case of WD~J2356$-$209 and WD~2251$-$070, there
remains a discrepancy between our models and the observations,
as there seems to be missing an absorption source in our models near 4500\,{\AA}.

The CaHe potentials and dipole moments can be obtained upon request
from T. Leininger and the \ion{Ca}{1} resonance line profiles from
N.~F.~Allard.

\acknowledgments

This work was supported in part by NSERC (Canada) and the Fund FRQNT (Qu\'ebec).

This work has made use of data from the European Space Agency (ESA) mission
{\it Gaia} (\url{https://www.cosmos.esa.int/gaia}), processed by the {\it Gaia}
Data Processing and Analysis Consortium (DPAC,
\url{https://www.cosmos.esa.int/web/gaia/dpac/consortium}). Funding for the DPAC
has been provided by national institutions, in particular the institutions
participating in the {\it Gaia} Multilateral Agreement.

\bibliographystyle{aasjournal}
\bibliography{references}

\end{document}